\documentclass[aps,prl,preprint,superscriptaddress,amsmath,amssymb]{revtex4-1}
\usepackage[utf8]{inputenc}
\usepackage{wasysym}
\usepackage{amsmath}
\usepackage{upgreek} % \textmu 
\newcommand{\cminv}{\, \mathrm{cm}^{-1}}
\newcommand{\microns}{\, \upmu \mathrm{m}}
\newcommand{\Plaser}{ P_{\mathrm{laser}} }

\newcommand{\mlpl}{\, \mathrm{mol/L}}
\newcommand{\mlpmin}{\, \mathrm{mL/min}}
\newcommand{\alwf}{\alpha_{\mathrm{w}}^{\mathrm{fiber}}}
\newcommand{\alef}{\alpha_{\mathrm{e}}^{\mathrm{fiber}}}
\newcommand{\aleff}{\alpha_{\mathrm{eff}}^{\mathrm{fiber}}}

\newcommand{\alcf}{\alpha_{\mathrm{c}}^{\mathrm{fiber}}}
\usepackage{graphicx}% Include figure files
\begin{document} 

%\preprint{}

\title{In-water chemical sensing by fiber-optic evanescent waves spectroscopy using mid-infrared quantum cascade lasers}



\author{Paul Chevalier}
%\thanks{}
\thanks{These authors contributed equally to this work.}
\affiliation{Harvard John A. Paulson School of Engineering and Applied Sciences, Harvard University, Cambridge, MA 02138 USA}

\author{Marco Piccardo}
%\thanks{}
\thanks{These authors contributed equally to this work.}
\affiliation{Harvard John A. Paulson School of Engineering and Applied Sciences, Harvard University, Cambridge, MA 02138 USA}

\author{Guy-Mael de Naurois}
%\thanks{}
\affiliation{Harvard John A. Paulson School of Engineering and Applied Sciences, Harvard University, Cambridge, MA 02138 USA}

\author{Ilan Gabay}
%\thanks{}
\affiliation{Harvard John A. Paulson School of Engineering and Applied Sciences, Harvard University, Cambridge, MA 02138 USA}
\affiliation{School of Physics and Astronomy, Sackler Faculty of Exact Sciences, Tel Aviv University, Tel Aviv, Israel}

\author{Abraham Katzir}
%\thanks{}
\affiliation{School of Physics and Astronomy, Sackler Faculty of Exact Sciences, Tel Aviv University, Tel Aviv, Israel}

\author{Federico Capasso}
\email[]{capasso@seas.harvard.edu}
%\thanks{}
\affiliation{Harvard John A. Paulson School of Engineering and Applied Sciences, Harvard University, Cambridge, MA 02138 USA}


\date{\today}

\begin{abstract}
The ability of detecting threatening chemicals diluted in water is an important safety requirement for drinking water systems. An apparatus for in-water chemical sensing based on the absorption of evanescent waves generated by a quantum cascade laser array and propagating inside a silver halide optical fiber immersed into water is demonstrated. We present a theoretical analysis of the sensitivity of the system and experimentally characterize its real-time response and spectroscopic detection for injection of a sample chemical (ethanol) in a tube containing water.
\end{abstract}

\pacs{}
%\keywords{}

%%\maketitle must follow title, authors, abstract, \pacs, and \keywords
\maketitle

\section{Introduction}

Water is polluted all over the world by highly toxic chemicals that are poured directly into it or into the ground by agriculture, industry, military and other sources. Since the toxicity of chemicals generated by these activities depends on their nature and ranges from highly toxic (e.g. organophosphorous pesticides or herbicides, such as Parathion, Diazinon or DDVP) to moderately toxic (e.g. halogenated hydrocarbons or ammonium perchlorate), it is not only necessary to detect their concentration but also to identify them inside a water system. A useful monitoring system should facilitate online monitoring in remote locations (e.g. inside water pipes or reservoirs), be sensitive, selective, robust, affordable and easy to operate. 

Mid-infrared (mid-IR) spectroscopy is a powerful tool to identify organic molecules as most of them exhibit molecular fingerprints due to  roto-vibrational modes absorbing in the mid-IR range~\cite{lin1991handbook}. Quantum cascade lasers (QCLs)~\cite{faist1994quantum} have demonstrated their ability as light sources for the spectroscopic detection of chemicals in the atmosphere reaching the part-per-billion sensitivity with the proper absorption cells~\cite{curl2010quantum}. Previous experiments were performed with such lasers operating in the mid-IR range for chemical sensing using a direct absorption method to detect phosphates~\cite{lendl2000mid}, adenine and xanthosine ~\cite{kolhed2002assessment}, glucose and fructose~\cite{edelmann2001towards}, or carbon dioxide~\cite{schaden2004direct} in aqueous samples by using a specific flow cell and a propagation path of the order of $100 \microns$. This short optical path is constrained by the large absorption coefficient of water~\cite{hale1973optical}, being about $500 \cminv$ (i.e. $20 \microns$ attenuation length) in the spectral range from $8 \microns$ to $10 \microns$, resulting in unpractical measurements using standard macroscopic transmission cells. On the other hand, the use of miniaturized cells would hamper the monitoring of large volumes of water. Measurement systems based on fiber-optic evanescent wave spectroscopy (FEWS) have been previously reported, either at visible wavelengths by using a surface plasmon resonance based sensor to detect pesticides~\cite{chand2007surface}, or in the infrared by using a silver halide fiber to measure in-water absorption of chemicals over a distance of few millimeters~\cite{Raichlin:08}. The latter system is based on attenuated total reflectance method allowing to measure the absorption of samples next to the fiber and relies on the high transparency~\cite{mizaikoff2003mid} in the mid-IR range, non-toxicity and non-hygroscopy of silver halide fibers. 
By combining QCLs with silver halide optical fibers, droplets of specific chemicals in liquid phase were directly detected~\cite{chen2005silver} by measuring the transmitted light spectrum, with a total immersed length below $25 \, \mathrm{mm}$.

Here, we present a sensing system for the \textit{in-situ} and real-time monitoring of chemical substances dispersed in water based on FEWS. The system integrates a multi-wavelength QCL array, a 80 cm-long silver halide fiber (AgClBr) contained in a PVC tube, and a fast mercury-cadmium-telluride (MCT) infrared detector. As the fiber is core-only, the evanescent tail of the electric field overlaps with the chemical diluted in the water (or any other fluid such as air) and the variations of the transmitted intensity can be detected on the infrared detector. We first explore the sensitivity of such a system with a theoretical analysis, then we demonstrate in-water transmission of a mid-IR evanescent wave and the detection of traces of ethanol. This chemical is chosen as a sample chemical for the safety of the experimental protocol but its absorption properties in the mid-IR range can be considered as representative of the class of highly toxic molecules that we are targeting in this study. Two configurations were used for the experimental study: a time-resolved transmission measurement at a single frequency that allows to characterize the time response of the system, and a static multi-wavelength study aiming at the spectroscopic identification of the molecular fingerprint of the chemical at different concentrations.

\begin{figure*}[ht]
\centering
\includegraphics[width=0.8\textwidth]{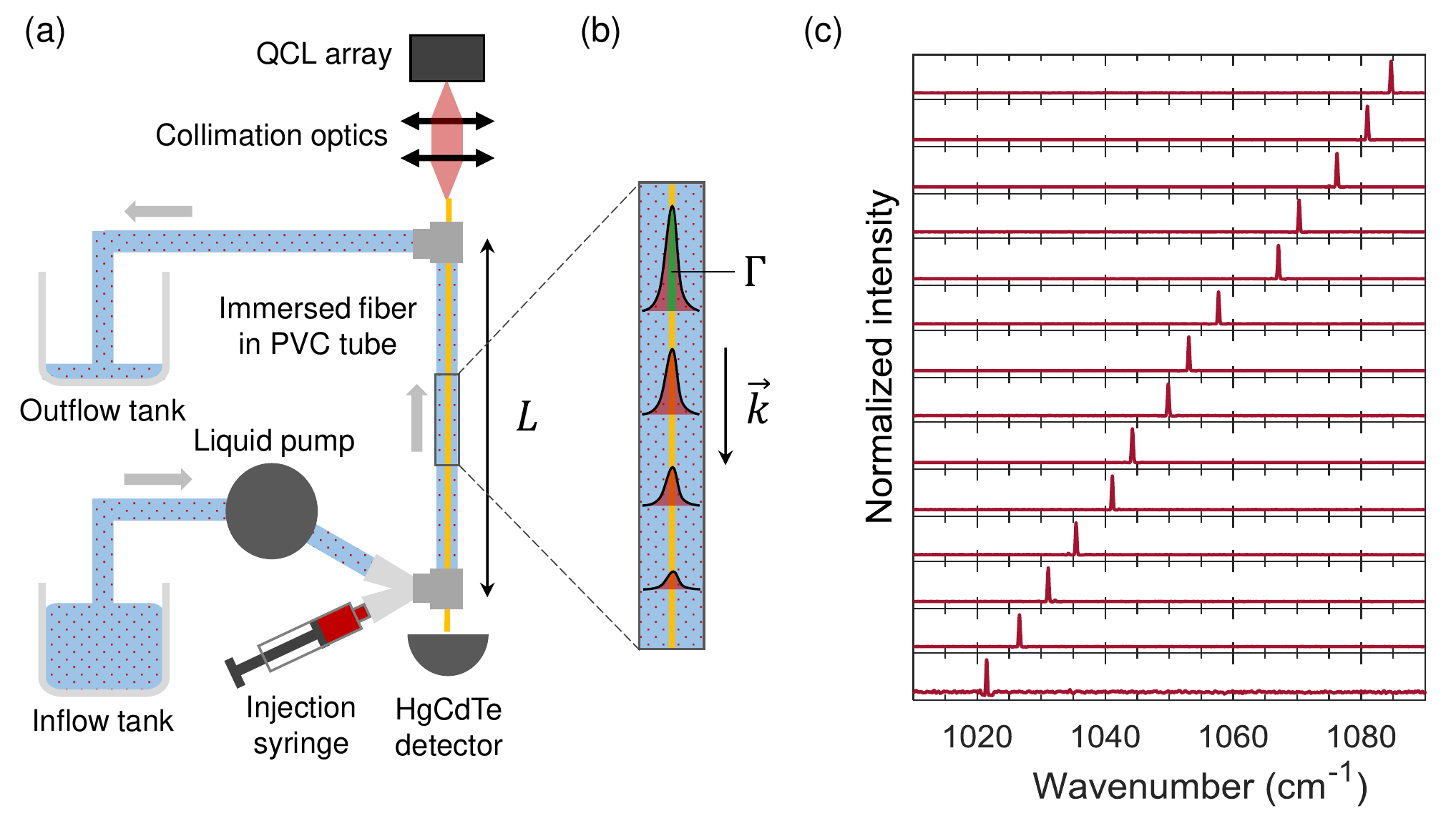}
  \caption{\label{fig:fig1}
  (a) Schematic of the fiber-optic system for in-water chemical sensing. A silver halide fiber lies in a PVC tube (inner diameter $6.4 \, \mathrm{mm}$, length $80 \, \mathrm{cm}$.) Water flows between two tanks propelled by a peristaltic pump.
  (b) Principle of fiber-optic evanescent wave sensing: as the light propagates along the immersed fiber (wavevector $\vec{k}$) with a confinement factor $\Gamma$, the evanescent tail overlaps with the liquid causing optical absorption.
  (c) Series of emission spectra produced by the different elements of the QCL array showing monochromatic emission from $1020 \cminv$ to $1085 \cminv$.
}
\end{figure*}

\section{Measurement setup}

A schematic of the measurement set-up is shown in Fig.~1(a): a multimode AgClBr fiber ($\diameter=0.9 \, \mathrm{mm}$, $L=80 \, \mathrm{cm}$) is fed inside a clear PVC tube (capacity 30 mL), with tee-shaped connectors at both ends. One port of the tees provides the interface for the fiber with the outside of the tube where light is coupled in/out of the fiber. Epoxy glue is used to guarantee a water-proof interface. The other port of the tees allows the fluid to flow in/out of the tube upon the propulsion generated by a peristaltic pump.

The mid-IR light source of the system consists of a QCL array based on the master-oscillator power-amplifier array geometry~\cite{rauter2013high}. The array is composed of 14 devices comprising two individual electrical sections: a power amplifier, and a distributed feedback (DFB) section.  The active region of this device is a GaInAs/AlInAs broadband bound-to-continuum heterostructure grown lattice-matched on a conducting InP substrate by organometallic vapor phase epitaxy (OMVPE). The grating period of the DFB section is varied for each device in order to obtain single-mode operation at specific wavelengths between $9.2$ and  $9.8 \microns$. 
The QCL array is operated in pulsed mode with a repetition rate of $10 \, \mathrm{kHz}$ and a pulse width of $80 \, \mathrm{ns}$. The emission spectrum of each device is measured using a Bruker 70 Fourier transform infrared spectrometer (cf. Fig.~1(c)).

The optical transmission of the fiber in the mid-IR is monitored by measuring the transmitted light generated from the different devices of the QCL array by using a thermoelectrically cooled mercury-cadmium-telluride (MCT) detector (Vigo PVI-4TE-10.6). The light pulses measured by the MCT detector are averaged using a box-car averager (Stanford Research Systems SR280) synchronized with the pulse generator and averaging over 300 pulses. The output signal of the box-car averager is acquired using a DAQ card (National Instruments BNC 2110).

The principle of FEWS is illustrated in Fig.~1(b): as the light generated by the laser propagates along the core-only fiber, the evanescent tails of the modes supported by the fiber overlap with the surrounding fluid and lead to a wavelength-dependent absorption of light in the fluid. As a consequence, the intensity decreases in the light wave along the whole fiber length. The efficiency of the evanescent wave absorption mechanism is determined by the fiber geometry (e.g. shape and size)~\cite{Raichlin:08}.

\section{Results and discussion}

\subsection{System analysis}

The performance of the sensing system is limited by a compromise that depends on the fiber length $L$, the laser power $\Plaser$, the noise level of the detector $N_0$ and the power coupling efficiency in and out of the fiber $\eta$. 
The bandwidth of the system for which the noise level is defined corresponds to the integration constant of the box-car averager used. This setting (300 samples at 10 kHz rate) is kept constant in all experiments leading to a bandwidth of approximately $30 \, \mathrm{Hz}$. This specific value was chosen to ensure fast response (within 100 ms) for the real-time study of the optical transmission.

The first constraint on the spectroscopy set-up requires that a transmitted signal larger than $N_0$ must be detected in presence of pure water, i.e. in absence of the chemical, which translates into the following relation:

\begin{equation}
P_{\mathrm{det}}^w =  \eta \Plaser  e^{-\alwf L} > N_0
\label{eq:rule1}
\end{equation}

where $\alwf$ is the effective absorption coefficient of water for light propagating in the fiber and $P_{\mathrm{det}}^w$ is the power measured by the detector when the tube is filled with pure water.

Eq.~\ref{eq:rule1} implies that the fiber length should be smaller than a critical length $L_c$:
 
\begin{equation}
L < \frac{1}{\alwf} \log \left (  \frac{\eta \Plaser}{N_0} \right ) =L_c
\label{eq:rule1l}
\end{equation}

The chemical to be detected absorbs light with an effective absorption coefficient $\alcf$ for the wave propagating in the fiber that follows the Beer-Lambert law and is thus proportional to its concentration. By using $\alwf$, $\alcf$ and the volumic concentration of the in-water dispersed chemical introduced in the fiber ($x$), the effective absorption coefficient due to the chemical dispersed in water is $\aleff(x)= x \alcf +(1-x) \alwf$. The power transmitted by the fiber when the tube is filled with water containing a volumic fraction $x$ of the target chemical, and measured by the detector $P_{\mathrm{det}}^c(x)$ is

\begin{equation}
P_{\mathrm{det}}^c(x) = \eta \Plaser e^{-\alwf L - x (\alcf-\alwf) L}
\label{eq:powerdetc}
\end{equation}

In order to successfully detect the target chemical, the change in transmission due to the chemical should be measurable, meaning that the difference in the power measured by the detector among the two previous situations must be larger than $N_0$.

\begin{multline}
P_{\mathrm{det}}^w - P_{\mathrm{det}}^c = 
\eta \Plaser e^{-\alwf L}\times \\
[ 1- e^{-x (\alcf-\alwf) L}]   > N_0
\label{eq:rule2}
\end{multline}

Combining Eq.~\ref{eq:rule1} and Eq.~\ref{eq:rule2} we can extract the minimum volumic fraction $x_{\mathrm{min}}$ that can be detected by the system:

\begin{equation}
x_{\mathrm{min}} = - \frac{1}{L(\alcf-\alwf)} \log \left ( 1 -  \frac{N_0}{P_{\mathrm{det}}^w} \right )
\label{eq:rule2l}
\end{equation}

%within at least 2 orders of magnitude
Assuming the condition stated in Eq.~\ref{eq:rule1} is strongly fulfilled  (i.e. $ P_{\mathrm{det}}^w  \gg N_0$ ), Eq.~\ref{eq:rule2l} can then be approximated by:

\begin{equation}
x_{\mathrm{min}}(L) \approx  \frac{e^{\alwf L}}{L(\alcf-\alwf)}  \frac{N_0}{\eta \Plaser}
\label{eq:cminapp}
\end{equation}

By taking the derivative of the Eq.~\ref{eq:cminapp} one can show that this expression is minimal for a fiber length given by $L_{\mathrm{opt}}=1/\alwf$, and the absolute minimum fraction of the chemical that can be detected with our system assuming all other parameters fixed becomes

\begin{equation}
x_{\mathrm{min}}(L_{\mathrm{opt}}) =  \frac{e~\alwf}{(\alcf-\alwf)}  \frac{N_0}{\eta \Plaser}
\label{eq:cminopt}
\end{equation}

\begin{figure}[htp]
\centering
\includegraphics[width=0.45\textwidth]{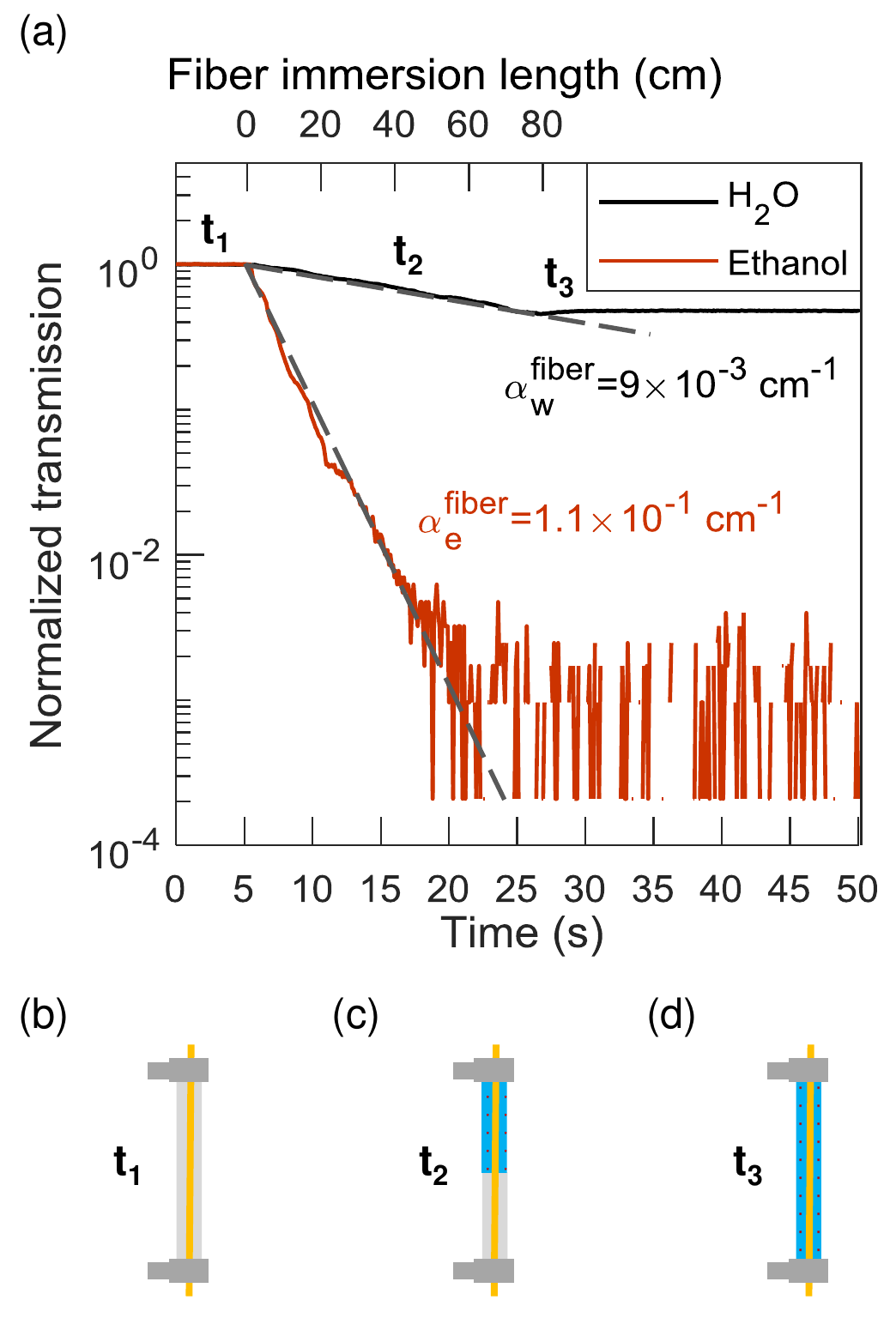}
\caption{\label{fig:fig2}
(a) 
The different steps of this procedure are the following:
(b) At $t = t_1$, the tube is empty and the fiber transmission is the highest.
(c) At $t=t_2$, the tube is partially filled and the transmission decreases.
(d) At $t = t_3$, the tube is completely filled and the transmission is the smallest.
}
\end{figure}

In order to determine the effective absorption coefficient of the fiber immersed in water or ethanol we performed a time-resolved measurement of the fiber transmission as the tube (initially purged with nitrogen) is progressively filled with the liquid. The transmission of the fiber at a wavenumber of $1044\cminv$, close to an ethanol absorption peak, is measured as a function of time as shown in Fig.~2 for ethanol and de-ionized water. During these measurements the tube is progressively filled with a regular pump flow of $60 \mlpmin$ preventing the creation of air pockets. We can thus linearly map the time with the length of fiber immersed in the liquid (see secondary x-axis in Fig.~2). Fitting the exponential decay of the transmission signal as a function of the immersed fiber length we extract $\alwf = 9 \times 10^{-3}  \cminv$ and $\alef = 1.1 \times 10^{-1}  \cminv$.
We also observe that when the tube is filled with pure ethanol, the transmission reaches the noise level at $t\approx20$ s. By taking the standard deviation of the noise signal we can determine the ratio of the noise level over the transmitted power as $N_0/(\eta \Plaser) = 2 \times 10^{-3}$. 

With the parameters extracted from this experiment we can determine using Eq.~\ref{eq:cminopt} the maximum sensitivity of the system. Considering the optimal fiber length $L_{\mathrm{opt}}= 111 \, \mathrm{cm}$ (close to that of our system: 80 cm), the minimum concentration that could be detected by the system is $x_{\mathrm{min}} \approx 0.048 \% $.

The system performance can be further increased by using more powerful lasers, more sensitive detectors, a larger integration time or flattened fibers to increase the overlap with the surrounding liquid~\cite{Raichlin:08}. 
%what else to say

%overlap factor is the same for 
%minimum sensitivity of the system: OK
%find other chemical and compute the min concentration

\begin{figure}[htp]
\centering
\includegraphics[width=0.43\textwidth]{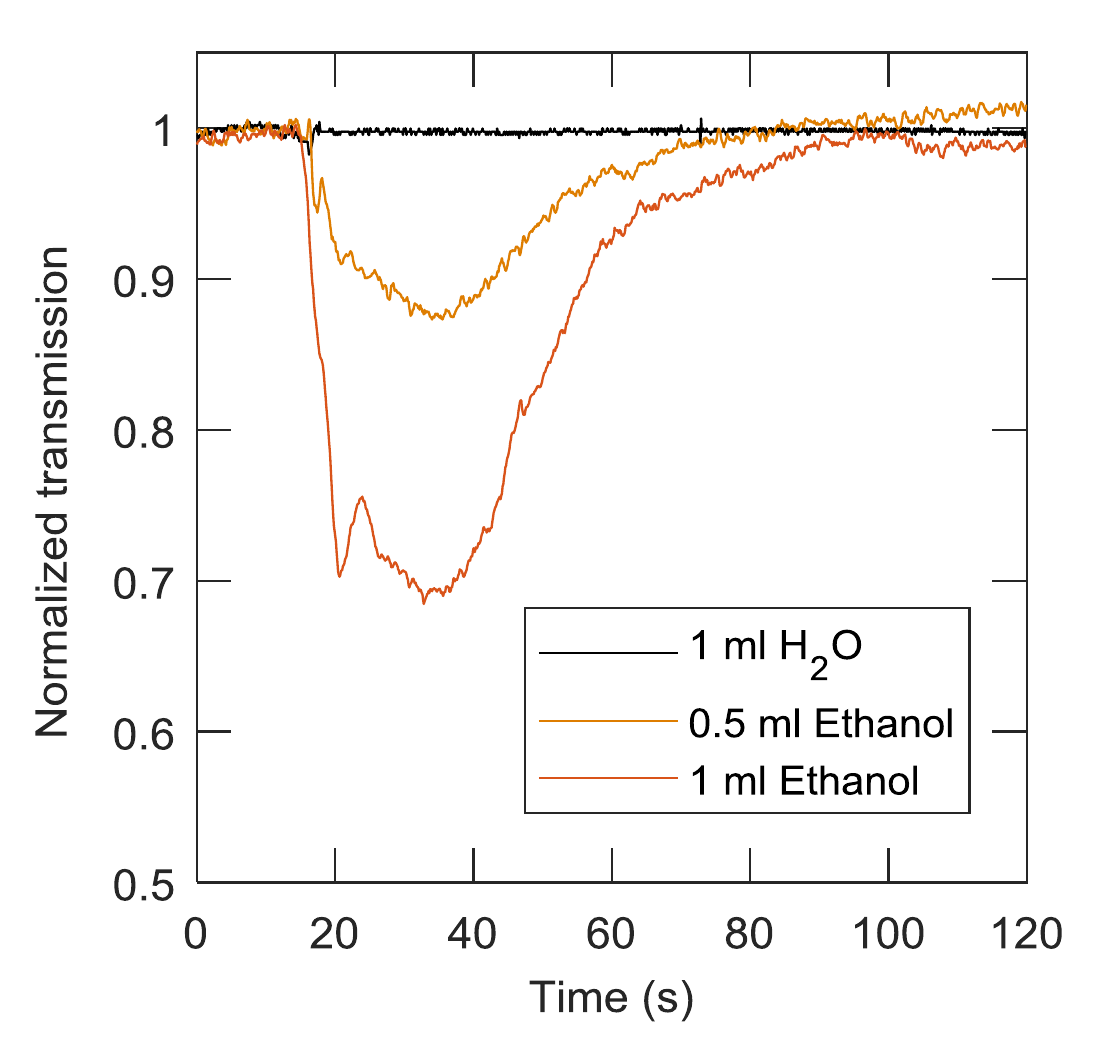}
  \caption{\label{fig:fig3}
Real-time monitoring of the fiber transmission when injecting a given volume of water or ethanol in the water flow. The laser wavenumber is $1050 \cminv$. 
}
\end{figure}

\subsection{Real-time monitoring}

In a second study, while a continuous flow of de-ionized water goes through the tube we inject using a Y-shaped connector (see Fig.~\ref{fig:fig1}a) a small volume of ethanol in the flow, either 0.5 or 1 mL, and simultaneously monitor the change in the fiber transmission at the wavenumber $ 1050 \cminv$, close to an absorption peak of ethanol. The injection occurs between $t=15 \, \mathrm{s}$ and $t=17 \, \mathrm{s}$ (see Fig.~3). We observe a rapid drop in transmission as the chemical comes in contact with the fiber. After approximately 1 min the ethanol diffuses (due to both the water flow and the physical diffusion) and the transmission increases back towards the previous level. For both injected volumes a sizeable drop in transmission (i.e. $>5\%$) is observed within the injection time. In a control experiment we inject 1 mL of water in the flow and no significant change in transmission occurs as the composition of the fluid surrounding the immersed fiber is unchanged. This result shows that injecting ethanol in the water flow leads to an immediate detection by a steep decrease in light transmission demonstrating the effectiveness of the system for real-time monitoring applications.

\subsection{Spectroscopic study}

\begin{figure}[htp]
\centering
\includegraphics[width=0.40\textwidth]{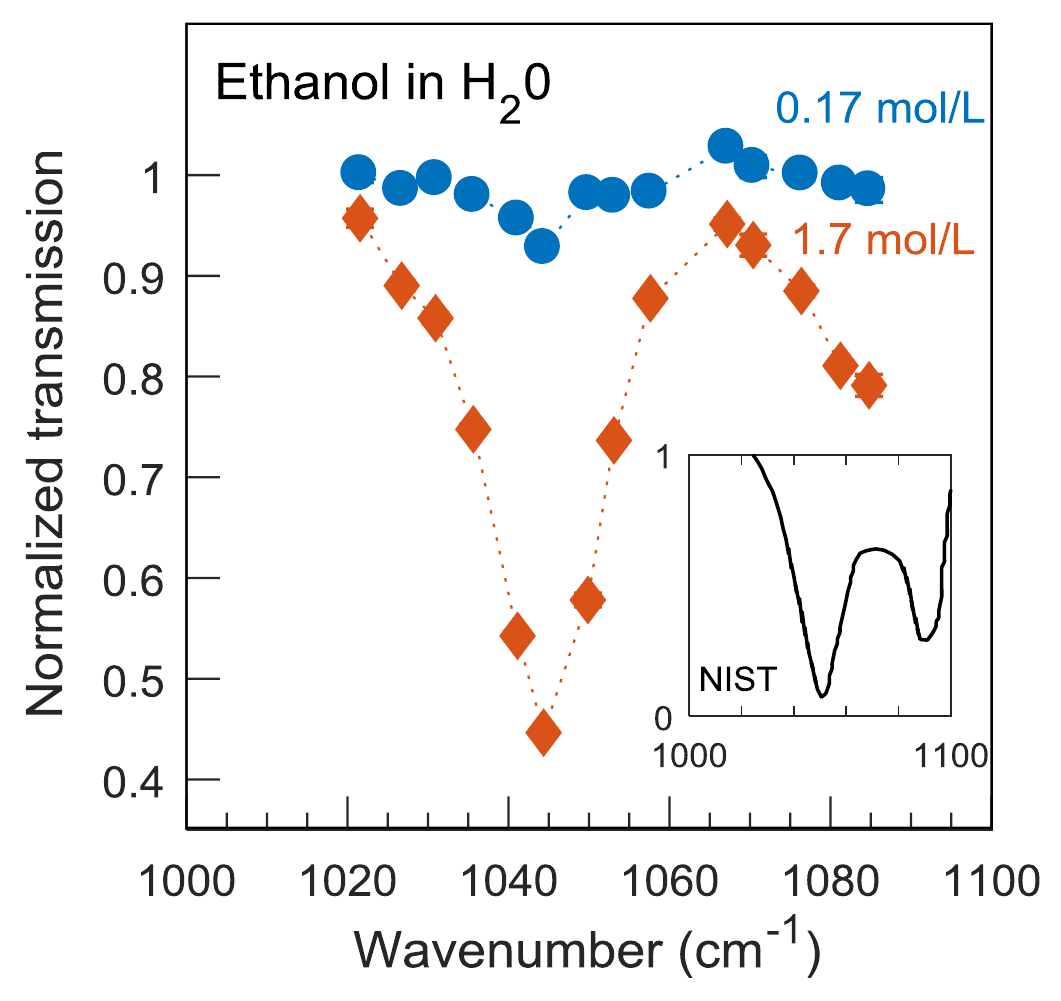}
\caption{\label{fig:fig4}
Spectroscopic study of the fiber transmission for different wavelengths in presence of various concentration of ethanol in the tube: $0.17 \mlpl$ in blue circles and $1.7 \mlpl$ in orange diamonds. The inset shows for comparison the mid-IR absorption spectrum of a solution of ethanol diluted in carbon tetrachloride (NIST, reproduced from Ref.~\cite{Plyler1952}).
}
\end{figure}
 
To demonstrate the full capacity of the system, we use the QCL array to measure the wavelength-dependent transmission of the fiber with different concentrations of ethanol in the tube. The resulting data, normalized by the transmission in pure water, are shown in Fig.~\ref{fig:fig4} for two different concentrations of ethanol, $0.17$ and $1.7 \mlpl$. The spectra measured by the FEWS system are in good agreement with the molecular fingerprint of ethanol in this spectral range (cf. inset of Fig.~\ref{fig:fig4}): a main pronounced transmission dip around $1045 \cminv$ and a local transmission maximum at $1070 \cminv$ are observed.

The study presented here focused on the detection of ethanol as a sample chemical diluted in water for the safety protocol of the laboratory environment. Harmful organophosphorus chemicals like pesticides (such as Parathion, Diazinon or DDVP) exhibit resonances in the spectral range considered in this manuscript with magnitudes even stronger than those featured by ethanol~\cite{tanner1996spectral}. Therefore the system presented here may be used to monitor drinking water sources, such as wells, in remote places where these pesticides have been used.

We also expect the proposed system to be suitable for the detection of harmful chemicals absorbing at different wavelengths in the mid-IR range~\cite{lin1991handbook} where the fiber is transparent and where QCL emission can be tailored~\cite{Yao2012}. Some examples of such chemicals accompanied by the wavenumber of their absorption peak are: cyanide ($2170 \cminv$), ammonium perchlorate ($1100 \cminv$) and trichloroethylene ($940 \cminv$).

For higher resolution spectroscopy, one may consider using QCLs with sampled grating distributed reflectors either for long wavelength~\cite{mansuripur2012widely} or shorter wavelength~\cite{kalchmair2015high} operation.

\section{Conclusions}

We presented a system providing fast detection of a chemical dispersed into water. A theoretical analysis provides insights into the main parameters relevant for the sensitivity of the apparatus and can guide future design optimization. We demonstrated both an alarm-type function by real-time monitoring of a single-wavelength transmission, and a spectroscopic function by measuring the transmission of the fiber at different wavelengths.
While the alarm-type response provides a prompt alert about the introduction of a foreign chemical into the water, the spectroscopic ability allows to discriminate and identify the nature of the chemical provided that its molecular fingerprint lies within the spectral range of the sensing system. Finally the AgClBr fibers used in this proof-of-concept are non-toxic and non-hygroscopic.
These feature makes this apparatus suitable for applications related to the online monitoring of drinking water systems. 

\section*{Acknowledgments}
The authors acknowledge Patrick Rauter, Stefan Menzel, Anish Goyal, Christine Wang, Antonio Sanchez and George Turner for fabricating the devices used in this paper.
This work was supported by the government of Israel, Ministry of Defense, Mission to the USA (Grant  no. 100313604) and  the  Defense  Threat  Reduction  Agency-Joint  Science  and  Technology  Office  for  Chemical  and  Biological  Defense  (Grant  no. HDTRA1-10-1-0031-DOD)

\end{document}